\pdfoutput=1
\documentclass[draftcls, onecolumn, 12pt]{IEEEtran}
\usepackage{amsmath,amsfonts,amssymb}
\usepackage{graphicx}
\usepackage{cite}
\usepackage{multirow}
\usepackage{algorithm}
\usepackage{algorithmic}
\usepackage{subfigure}

\IEEEoverridecommandlockouts

\begin{document}
\title{Detailed Steps of the Fourier-Motzkin Elimination}
\author{Xiaoli Xu, Yong Zeng, Yong Liang Guan and Tracey, Ho }
\maketitle
This file provide the detailed steps for obtaining the bounds on $R_1$, $R_2$ via the obtained results on $(R_{1c},R_{1p},R_{2c},R_{2p})$.  The rate pair $(R_{1c},R_{1p},R_{2c},R_{2p})$ is achievable if it satisfies following inequalities.
\begin{align}
R_{1c}&\leq \mathrm{rank}(U_{1}^TH_{11}V_{21})\label{eq:cp11}\\
R_{1p}&\leq m_1-r_{21}\label{eq:cp12}\\
R_{2c}&\leq r_{12}\label{eq:cp13}\\
R_{1c}+R_{1p}&\leq r_{11}\label{eq:cp14}\\
R_{1c}+R_{2c}&\leq r_{12}+\mathrm{rank}(U_{10}^TH_{11}V_{21})\label{eq:cp15}\\
R_{1p}+R_{2c} &\leq r_{12}+\mathrm{rank}(U_{10}^TH_{11}V_{20})\label{eq:cp16}\\
R_{1c}+R_{1p}+R_{2c}&\leq n_1\label{eq:cp17}\\
R_{2c}&\leq \mathrm{rank}(U_{2}^TH_{22}V_{11})\label{eq:cp21}\\
R_{2p}&\leq m_2-r_{12}\label{eq:cp22}\\
R_{1c}&\leq r_{21}\label{eq:cp23}\\
R_{2c}+R_{2p}&\leq r_{22}\label{eq:cp24}\\
R_{2c}+R_{1c}&\leq r_{21}+\mathrm{rank}(U_{20}^TH_{22}V_{11})\label{eq:cp25}\\
R_{2p}+R_{1c} &\leq r_{21}+\mathrm{rank}(U_{20}^TH_{22}V_{10})\label{eq:cp26}\\
R_{2c}+R_{2p}+R_{1c}&\leq n_2\label{eq:cp27}
\end{align}
where \eqref{eq:cp13} is implied by \eqref{eq:cp21} as $\mathrm{rank}(U_{2}^TH_{22}V_{11})\leq\mathrm{rank}(V_{11})=r_{12}$. Similarly, \eqref{eq:cp23} is implied by \eqref{eq:cp11}. Removing the redundant inequalities and substituting with $R_{1p}=R_1-R_{1c}$, $R_{2p}=R_2-R_{2c}$, we obtain
\begin{align}
R_{1c}&\leq \mathrm{rank}(U_{1}^TH_{11}V_{21})\label{eq:v11}\\
R_{1}-R_{1c}&\leq m_1-r_{21}\label{eq:v12}\\
R_1&\leq r_{11}\label{eq:v13}\\
R_{1c}+R_{2c}&\leq r_{12}+\mathrm{rank}(U_{10}^TH_{11}V_{21})\label{eq:v14}\\
R_{1}-R_{1c}+R_{2c} &\leq r_{12}+\mathrm{rank}(U_{10}^TH_{11}V_{20})\label{eq:v15}\\
R_1+R_{2c}&\leq n_1\label{eq:v16}\\
R_{2c}&\leq \mathrm{rank}(U_{2}^TH_{22}V_{11})\label{eq:v17}\\
R_{2}-R_{2c}&\leq m_2-r_{12}\label{eq:v18}\\
R_2&\leq r_{22}\label{eq:v19}\\
R_{2c}+R_{1c}&\leq r_{21}+\mathrm{rank}(U_{20}^TH_{22}V_{11})\label{eq:v110}\\
R_{2}-R_{2c}+R_{1c} &\leq r_{21}+\mathrm{rank}(U_{20}^TH_{22}V_{10})\label{eq:v111}\\
R_{2}+R_{1c}&\leq n_2\label{eq:v112}
\end{align}
To eliminate $R_{1c}$, we get five upper bounds and two lower bounds on $R_{1c}$
\begin{align*}
R_{1c}&\leq \mathrm{rank}(U_{1}^TH_{11}V_{21})\\
R_{1c}&\leq r_{12}+\mathrm{rank}(U_{10}^TH_{11}V_{21})-R_{2c}\\
R_{1c}&\leq r_{21}+\mathrm{rank}(U_{20}^TH_{22}V_{11})-R_{2c}\\
R_{1c} &\leq r_{21}+\mathrm{rank}(U_{20}^TH_{22}V_{10})-R_{2}+R_{2c}\\
R_{1c}&\leq n_2-R_{2}\\
R_{1c}&\geq R_{1}-m_1+r_{21}\\
R_{1c} &\geq R_{1}+R_{2c}-r_{12}-\mathrm{rank}(U_{10}^TH_{11}V_{20})
\end{align*}
Comparing the upper and lower bounds of $R_{1c}$ and copying five inequalities in \eqref{eq:v11}-\eqref{eq:v112} that do not involve $R_{1c}$, we obtain a new system of inequalities in $(R_{2c},R_1,R_2)$ give by
\begin{align}
R_1&\leq r_{11} \label{eq:v21}\\
R_1+R_{2c}&\leq n_1\label{eq:v22}\\
R_{2c}&\leq \mathrm{rank}(U_{2}^TH_{22}V_{11})\label{eq:v23}\\
R_{2}-R_{2c}&\leq m_2-r_{12}\label{eq:v24}\\
R_2&\leq r_{22}\label{eq:v25}\\
R_1&\leq \mathrm{rank}(U_{1}^TH_{11}V_{21})+m_1-r_{21}\label{eq:v26}\\
R_1+R_{2c}&\leq r_{12}+\mathrm{rank}(U_{10}^TH_{11}V_{21})+m_1-r_{21}\label{eq:v210}\\
R_1+R_{2c}&\leq \mathrm{rank}(U_{20}^TH_{22}V_{11})+m_1\label{eq:v27}\\
R_1+R_2-R_{2c}&\leq \mathrm{rank}(U_{20}^TH_{22}V_{10})+m_1\label{eq:v28}\\
R_1+R_2&\leq n_2+m_1-r_{21}\label{eq:v29}\\
R_1+R_{2c}&\leq \mathrm{rank}(U_{1}^TH_{11}V_{21})+\mathrm{rank}(U_{10}^TH_{11}V_{20})+r_{12}\label{eq:v211}\\
R_1+2R_{2c}&\leq 2r_{12}+\mathrm{rank}(U_{10}^TH_{11}V_{21})+\mathrm{rank}(U_{10}^TH_{11}V_{20})\label{eq:v215}\\
R_1+2R_{2c}&\leq r_{21}+\mathrm{rank}(U_{20}^TH_{22}V_{11})+\mathrm{rank}(U_{10}^TH_{11}V_{20})+r_{12}\label{eq:v212}\\
R_1+R_2 &\leq \mathrm{rank}(U_{20}^TH_{22}V_{10})+\mathrm{rank}(U_{10}^TH_{11}V_{20})+r_{12}+r_{21}\label{eq:v213}\\
R_1+R_2+R_{2c}&\leq n_2+r_{12}+\mathrm{rank}(U_{10}^TH_{11}V_{20})\label{eq:v214}
\end{align}
where \eqref{eq:v26} is implied by \eqref{eq:v21} as $\mathrm{rank}(U_{1}^TH_{11}V_{21})=\mathrm{rank}(H_{11}V_{21})\geq r_{11}-m_1+r_{21}$; \eqref{eq:v210} is impled by \eqref{eq:v22} as $\mathrm{rank}(U_{10}^{T}H_{11}V_{21})\geq n_1-m_1+r_{21}-r_{12}$; \eqref{eq:v211} is impled by \eqref{eq:v22} as $\mathrm{rank}(U_{1}^TH_{11}V_{21})+\mathrm{rank}(U_{10}^TH_{11}V_{20})+r_{12}
=\mathrm{rank}(H_{11}V_{21})+\mathrm{rank}(U_{10}^TH_{11}V_{20})+r_{12}
\geq r_{11}+r_{21}-m_1+r_{12}+n_1+m_1-r_{11}-r_{12}-r_{21}=n_1$.

Now, we are ready to eliminate $R_{2c}$ from the remaining non-redundant inequalities. In total, six upper bounds and two lower bounds are obtained for $R_{2c}$.
\begin{align*}
R_{2c}&\leq n_1-R_1\\
R_{2c}&\leq \mathrm{rank}(U_{2}^TH_{22}V_{11})\\
R_{2c}&\leq \mathrm{rank}(U_{20}^TH_{22}V_{11})+m_1-R_1\\
2R_{2c}&\leq r_{21}+\mathrm{rank}(U_{20}^TH_{22}V_{11})+\mathrm{rank}(U_{10}^TH_{11}V_{20})+r_{12}-R_{1}\\
R_{2c}&\leq n_2+r_{12}+\mathrm{rank}(U_{10}^TH_{11}V_{20})-R_1-R_2\\
2R_{2c}&\leq 2r_{12}+\mathrm{rank}(U_{10}^TH_{11}V_{21})+\mathrm{rank}(U_{10}^TH_{11}V_{20})-R_{1}\\
R_{2c}&\geq R_{2}-m_2+r_{12}\\
R_{2c}&\geq R_1+R_2-\mathrm{rank}(U_{20}^TH_{22}V_{10})-m_1
\end{align*}
Comparing the lower and upper bounds of $R_{2c}$, we get 12 inequalities. Together with the inequalities in \eqref{eq:v21}-\eqref{eq:v215} that do not involve $R_{2c}$, we obtain a new system of inequalities in $(R_1,R_2)$ given by
\begin{align}
R_1&\leq r_{11} \label{eq:v31}\\
R_2&\leq r_{22}\label{eq:v32}\\
R_1+R_2&\leq n_2+m_1-r_{21}\label{eq:v33}\\
R_1+R_2 &\leq \mathrm{rank}(U_{20}^TH_{22}V_{10})+\mathrm{rank}(U_{10}^TH_{11}V_{20})+r_{12}+r_{21}\label{eq:v34}\\
R_1+R_2 &\leq n_1+m_2-r_{12}\label{eq:v35}\\
R_2 &\leq \mathrm{rank}(U_{2}^TH_{22}V_{11})+m_2-r_{12}\label{eq:v36}\\
R_1+R_2 &\leq \mathrm{rank}(U_{20}^TH_{22}V_{11})+m_1+m_2-r_{12}\label{eq:v310}\\
R_1+2R_2 &\leq r_{21}+\mathrm{rank}(U_{20}^TH_{22}V_{11})+\mathrm{rank}(U_{10}^TH_{11}V_{20})+2m_2-r_{12}\label{eq:v37}\\
R_1+2R_2 &\leq n_2+m_2+\mathrm{rank}(U_{10}^TH_{11}V_{20})\label{eq:v38}\\
R_1+2R_2 &\leq \mathrm{rank}(U_{10}^TH_{11}V_{21})+\mathrm{rank}(U_{10}^TH_{11}V_{20})+2m_2\label{eq:v39}\\
2R_1+R_2 &\leq n_1+\mathrm{rank}(U_{20}^TH_{22}V_{10})+m_1\label{eq:v311}\\
R_1+R_2 &\leq \mathrm{rank}(U_{2}^TH_{22}V_{11})+\mathrm{rank}(U_{20}^TH_{22}V_{10})+m_1\label{eq:v312}\\
2R_1+R_2 &\leq \mathrm{rank}(U_{20}^TH_{22}V_{11})+2m_1+\mathrm{rank}(U_{20}^TH_{22}V_{10})\label{eq:v316}\\
3R_1+2R_2 &\leq \mathrm{rank}(U_{20}^TH_{22}V_{11})+r_{12}+r_{21}+\mathrm{rank}(U_{10}^TH_{11}V_{20})+2\mathrm{rank}(U_{20}^TH_{22}V_{10})+2m_1\label{eq:v313}\\
2R_1+2R_2 &\leq n_2+r_{12}+\mathrm{rank}(U_{10}^TH_{11}V_{20})+\mathrm{rank}(U_{20}^TH_{22}V_{10})+m_1\label{eq:v314}\\
3R_1+2R_2 &\leq 2r_{12}+\mathrm{rank}(U_{10}^TH_{11}V_{21})+\mathrm{rank}(U_{10}^TH_{11}V_{20})+2\mathrm{rank}(U_{20}^TH_{22}V_{10})+2m_1\label{eq:v315}
\end{align}
Now, we are proceed to identifying the redundant inequalities.

\eqref{eq:v36} is implied by \eqref{eq:v32} as $\mathrm{rank}(U_{2}^TH_{22}V_{11})=\mathrm{rank}(H_{22}V_{11})\geq r_{22}+r_{12}-m_2$.

\eqref{eq:v37} is implied by \eqref{eq:v38} as $\mathrm{rank}(U_{20}^{T}H_{22}V_{11})\geq n_2-m_2+r_{12}-r_{21}$.

\eqref{eq:v39} is implied by \eqref{eq:v32} and \eqref{eq:v35} as
\begin{align*}
\mathrm{rank}(U_{10}^TH_{11}V_{21})+\mathrm{rank}(U_{10}^TH_{11}V_{20})+2m_2
&\geq \mathrm{rank}(U_{10}^TH_{11})+2m_2\\
&\geq n_1-r_{12}+2m_2\geq n_1+m_2-r_{12}+r_{22}
\end{align*}

\eqref{eq:v310} is implied by \eqref{eq:v33} as $\mathrm{rank}(U_{20}^{T}H_{22}V_{11})\geq n_2-m_2+r_{12}-r_{21}$.

\eqref{eq:v312} is implied by \eqref{eq:v33} as $\mathrm{rank}(U_{2}^TH_{22}V_{11})+\mathrm{rank}(U_{20}^TH_{22}V_{10})+m_1
=\mathrm{rank}(H_{22}V_{11})+\mathrm{rank}(U_{20}^TH_{22}V_{10})+m_1\geq n_2-r_{21}+m_1$.

\eqref{eq:v313} can be obtained by summing up \eqref{eq:v34} and \eqref{eq:v316}.

\eqref{eq:v314} can be obtained by summing up \eqref{eq:v33} and \eqref{eq:v34}.

\eqref{eq:v315} is implied by \eqref{eq:v34} and \eqref{eq:v311}.

\eqref{eq:v316} is implied by \eqref{eq:v31} and \eqref{eq:v33} as
\begin{align*}
\mathrm{rank}(U_{20}^TH_{22}V_{11})+2m_1+\mathrm{rank}(U_{20}^TH_{22}V_{10})
&\geq \mathrm{rank}(U_{20}^TH_{22})+2m_1\\
&=n_2-r_{21}+2m_1\geq n_2-r_{21}+m_1+r_{11}
\end{align*}

After removing all the redundant inequalities, we obtain the non-trivial set of inequalities.
\begin{align}
R_1&\leq r_{11}\label{eq:bound1}\\
R_2&\leq r_{22}\label{eq:bound2}\\
R_1+R_2&\leq n_{2}+m_{1}-r_{21}\label{eq:bound3}\\
R_1+R_2&\leq n_1+m_2-r_{12}\label{eq:bound4}\\
R_1+R_2&\leq \mathrm{rank}(U_{20}^TH_{22}V_{10})+\mathrm{rank}(U_{10}^TH_{11}V_{20})+r_{21}+r_{12}\label{eq:bound5}\\
2R_1+R_2&\leq \mathrm{rank}(U_{20}^TH_{22}V_{10})+m_1+n_1\label{eq:bound6}\\
R_1+2R_2&\leq \mathrm{rank}(U_{10}^TH_{11}V_{20})+m_2+n_2\label{eq:bound7}
\end{align}

\appendices
\section{List of Facts}
List of facts that are useful for in Fourier-Motzkin Elimination
\begin{enumerate}
\item{$\mathrm{rank}(U_{10}^TH_{11})=n_1-r_{12}$}
\item{$\mathrm{rank}(U_{20}^TH_{22})=n_2-r_{21}$}
\item{$\mathrm{rank}(H_{11}V_{20})=m_1-r_{21}$}
\item{$\mathrm{rank}(H_{22}V_{10})=m_2-r_{12}$}
\item{$\mathrm{rank}(H_{11}V_{21})\geq r_{11}+r_{21}-m_1$}
\item{$\mathrm{rank}(H_{22}V_{11})\geq r_{22}+r_{12}-m_2$}
\item{$\mathrm{rank}(U_{10}^{T}H_{11}V_{21})\geq n_1-m_1+r_{21}-r_{12}$}
\item{$\mathrm{rank}(U_{20}^{T}H_{22}V_{11})\geq n_2-m_2+r_{12}-r_{21}$}
\item{$\mathrm{rank}(U_{10}^{T}H_{11}V_{20})\geq n_1+m_1-r_{11}-r_{12}-r_{21}$}
\item{$\mathrm{rank}(U_{20}^{T}H_{22}V_{10})\geq n_2+m_2-r_{22}-r_{12}-r_{21}$}
\end{enumerate}
\begin{proof}
Fact 1 can be proved as
\begin{align*}
\mathrm{rank}(U_{10}^TH_{11})&=\mathrm{rank}(H_{11})-\mathrm{dim}(\mathcal {N}(U_{10}^T)\cap\mathcal{R}(H_{11}))\\
&=r_{11}-\mathrm{dim}(\mathcal {R}(H_{12})\cap\mathcal{R}(H_{11}))\\
&=r_{11}-(r_{11}+r_{12}-\mathrm{rank}([\begin{matrix}H_{11} & H_{12}\end{matrix}]))\\
&=\mathrm{rank}([\begin{matrix}H_{11} & H_{12}\end{matrix}])-r_{12}\\
&\overset{(a)}{=}n_{1}-r_{12}
\end{align*}
where (a) follows as $\left[\begin{matrix}H_{11} & H_{12}\end{matrix}\right]$ is full row rank. By symmetry, Fact 2 can be proved.

To prove Fact 3
\begin{align*}
\mathrm{rank}(H_{11}V_{20})
&=\mathrm{rank}(V_{20})-\mathrm{dim}(\mathcal{N}(H_{11})\cap\mathcal{R}(V_{20}))\\
&=m_1-r_{21}-\mathrm{dim}(\mathcal{N}(H_{11})\cap\mathcal{N}(H_{21}))\\
&\overset{(b)}{=}m_1-r_{21}
\end{align*}
where (b) follows as $\left[\begin{matrix}H_{11}\\H_{21}\end{matrix}\right]$ is full column rank. By symmetry, Fact 4 can be proved.

Fact 5 can be proved as $\mathrm{rank}(H_{11})=\mathrm{rank}\left(H_{11}\left[\begin{matrix}V_{21}& V_{20}\end{matrix}\right]\right)\leq\mathrm{rank}(H_{11}V_{21})+\mathrm{rank}(H_{11}V_{20})$. Since $\mathrm{rank}(H_{11}V_{20})=m_1-r_{21}$, $\mathrm{rank}(H_{11}V_{21})\geq r_{11}+r_{21}-m_1$. Fact 6 follows similarly.

Fact 7 can be proved using Frobenius inequality,
\begin{align*}
\mathrm{rank}(U_{10}^{T}H_{11}V_{21})&\geq\mathrm{rank}(U_{10}^TH_{11})+\mathrm{rank}(H_{11}V_{21})-\mathrm{rank}(H_{11})\\
&\geq n_1-r_{12}+r_{11}+r_{21}-m_1-r_{11}\\
&=n_1-m_1+r_{21}-r_{12}
\end{align*}
Fact 8 follows similarly.

Fact 9 is also proved using Frobenius inequality,
\begin{align*}
\mathrm{rank}(U_{10}^{T}H_{11}V_{20})&\geq\mathrm{rank}(U_{10}^{T}H_{11})+\mathrm{rank}(H_{11}V_{20})-\mathrm{rank}(H_{11})\\
&=n_1-r_{12}+m_1-r_{21}-r_{22}
\end{align*}
Fact 10 follows similarly.
\end{proof}

\end{document}